\documentclass{imsart} 

\usepackage{graphicx}
\usepackage{dcolumn}
\usepackage{bm}
\usepackage{amsthm}
\usepackage{amsmath}
\usepackage{natbib}
\usepackage[colorlinks,citecolor=blue,urlcolor=blue,filecolor=blue,backref=page]{hyperref}
\usepackage{graphicx}
\usepackage{natbib}
\usepackage{xspace}

\usepackage[english]{babel}
\usepackage[T1]{fontenc}
\usepackage[utf8]{inputenc}
\usepackage[scaled=1.04]{biolinum}

\usepackage{fourier}
\usepackage[scaled=0.83]{beramono}

\newcommand{\like}{\mathcal{L}}
\newcommand{\pvalue}{\textit{p}-value\xspace}
\newcommand{\given}{\, \middle| \,}
\newcommand{\pg}[2]{p\left(#1 \given #2 \right)}
\newcommand{\intd}{\text{d}}
\newcommand{\normal}{\mathcal{N}}

\newcommand{\refeq}{eq.~\ref}
\newcommand{\refsec}{sec.~\ref}
\newcommand{\refcite}{\cite}
\newcommand{\reffig}{fig.~\ref}
\newcommand{\pcite}[1]{(\cite{#1})}

\startlocaldefs
\numberwithin{equation}{section}
\theoremstyle{plain}

\endlocaldefs

\begin{document}

\begin{frontmatter}
\title{Objective Bayesian approach to the Jeffreys-Lindley paradox}
\runtitle{Objective approach to Lindley's paradox}

\begin{aug}

\author{\fnms{Andrew} \snm{Fowlie}\thanksref{addr1}\ead[label=e1]{andrew.j.fowlie@njnu.edu.cn}}

\runauthor{A.~Fowlie}

\address[addr1]{Department of Physics and Institute of Theoretical Physics, Nanjing Normal University,
  Nanjing, Jiangsu 210023, China,
  \printead{e1} 
}

\end{aug}

\begin{abstract}
We consider the Jeffreys-Lindley paradox from an objective Bayesian perspective by attempting to find priors representing complete indifference to sample size in the problem. This means that we ensure that the prior for the unknown mean and the prior predictive for the $t$-statistic are independent of the sample size. If successful, this would lead to Bayesian model comparison that was independent of sample size and ameliorate the paradox. Unfortunately, it leads to an improper scale-invariant prior for the unknown mean. We show, however, that a truncated scale-invariant prior delays the dependence on sample size, which could be practically significant. Lastly, we shed light on the paradox by relating it to the fact that the scale-invariant prior is improper.
\end{abstract}



\end{frontmatter}

\section{Introduction}

The Jeffreys-Lindley (JL) paradox~\pcite{Jeffreys:1939xee,10.2307/2333251} exposes a conflict between Bayesian and frequentist approaches to hypothesis testing that grows as the sample size increases. Whilst the mathematics of the paradox are clear, its repercussions and meaning continue to be disputed to this day. In this note, we emphasise that an objective Bayesian approach partially ameliorates the paradox. This was previously explored in passing in \refcite{Fowlie:2019ydo}.

To establish our notation, we briefly review the paradox in \refsec{sec:review} (see e.g.,~\refcite{doi:10.1080/01621459.1982.10477809,doi:10.1086/673730,doi:10.1086/675729,Cousins:2013hry} for further introduction and discussion). In \refsec{sec:result} we construct priors representing complete indifference to the sample size. We show that in this case the dependence of the Bayes factor on the sample size and thus the JL paradox vanish. We comment further on this result and conclude in \refsec{sec:conclusions}.

\section{The JL paradox}\label{sec:review}

The problem under consideration is as follows. We collect a sample of size $n$ from a normal distribution, $\normal(\mu, \sigma^2)$. We suppose that $\sigma$ and $n$ are known and that we want to test whether $\mu = 0$ ($H_0$) or $\mu \neq 0$ ($H_1$). In both frequentist and Bayesian approaches, we may summarise the data by the sufficient $t$-statistic
\begin{equation}\label{eq:t}
t = \frac{\sqrt{n} \bar x}{\sigma},
\end{equation}
where $\bar x$ is the sample mean. The likelihood may be written as
\begin{equation}\label{eq:like}
\like = \frac{1}{\sqrt{2\pi}} e^{-\tfrac12 (t - t_0)^2},
\end{equation}
where  $t_0 = \sqrt{n} \mu / \sigma$. In this form, it depends on $n$ and $\mu$ only through the combination $\sqrt{n} \mu$.

The $t$-statistic is distributed as a standard normal variable, $\normal(0, 1)$, under the null hypothesis in which $\mu = 0$. In the frequentist approach, we may thus compute a \pvalue,
\begin{equation}
\text{\pvalue} = 2 \left(1  - \Phi(|t|)\right).
\end{equation}
where $\Phi$ is the cumulative distribution function of a standard normal distribution. We emphasise that the \pvalue depends only on $t$ and not on $n$.

In the Bayesian approach, on the other hand, we compute a Bayes factor (see e.g., \refcite{doi:10.1080/01621459.1995.10476572}) for the alternative versus the null hypothesis,
\begin{align}
B_{10} ={}& \frac{
              \pg{t}{H_1}
            }{
              \pg{t}{H_0}
            }\\
        ={}& \frac{
              \int \pg{t}{\mu} p(\mu) \, \intd \mu
            }{
              \pg{t}{H_0}
            }\\
       ={}& e^{\frac12 t^2} \int e^{-\tfrac12 (t - \sqrt{n}\mu/\sigma)^2} p(\mu) \, \intd \mu.\label{eq:bf2}
\end{align}
This required that we placed a prior, $p(\mu)$, on the unknown mean in the alternative model. By a Laplace approximation, we find that for any slowly varying prior,
\begin{align}
B_{10} \approx \sqrt{2\pi} e^{\frac12 t^2}  \frac{p(\hat\mu) \sigma}{\sqrt{n}},
\end{align}
where $\hat\mu = t \sigma / \sqrt{n}$ is the maximum of the exponential factor in the integrand.

We see that whilst the \pvalue depends only on the observed $t$, the Bayes factor depends on $t$ and the sample size $n$. For any $t$ for which we would reject the null hypothesis via the \pvalue, the Bayes factor could favour the null by an arbitrary factor by making $n$ sufficiently large. This is the JL paradox. It concerns changing $n$ and a fixed observation of $t$.

\section{Objective Bayesian approach}\label{sec:result}

In the JL paradox, $t$ was fixed whereas the sample size, $n$, was changed. Since the \pvalue is independent of $n$, for a fair comparison, let us attempt to construct a prior $p(\mu)$ representing complete indifference to $n$ in the problem. Specifically, as the problem concerns inferences about $\mu$ in light of an observation of $t$, the prior should be such that changing $n$ changes nothing about our prior knowledge of $t$ or $\mu$ in the alternative hypothesis. The null hypothesis is always indifferent to $n$, as the prior predictive $\pg{t}{H_0}$ is independent of $n$.

By construction, for such a prior, the prior predictive for $t$ should be independent of $n$, and therefore the evidence
$\pg{t}{H_1}$ and the Bayes factor are independent of $n$. The JL paradox would thus vanish for such priors, as the frequentist and Bayesian results would both be independent of $n$. To find such a prior, it suffices to demand that the prior for the unknown $\mu$, $p(\mu)$ and the subsequent prior for the unknown $t_0$, $q(t_0)$ are independent of $n$. By the Jacobian rule and the relation $t_0 = \sqrt{n} \mu / \sigma$, however, we require that
\begin{align}
p(\mu) = {}& p(\sigma / \sqrt{n} t_0) \\
       = {}& q(t_0) \frac{\sqrt{n}}{\sigma} = q(\sqrt{n} \mu / \sigma) \frac{\sqrt{n}}{\sigma}.
\end{align}
The solution is thus
\begin{equation}\label{eq:log_prior}
p(\mu) = \frac{c}{|\mu|},
\end{equation}
for constant $c$. This is the usual scale-invariant prior (see e.g., \refcite{consonni2018,doi:10.1080/01621459.1996.10477003} for further discussion of invariant priors and rules for finding priors), which is improper, reflecting the fact that complete indifference to scale is impossible.\footnote{Since $\sqrt{n} / \sigma$ is positive in our problem, we are in fact free to weight positive and negative $\mu$ differently, but we won't pursue that complication here.} Note, however, that the prior predictive for $t$, whilst independent of $n$, is not scale-invariant.

We may quickly verify that with this prior, the Bayes factor in \refeq{eq:bf2} may be written,
\begin{align}\label{eq:b10}
B_{10} = {}& c e^{\tfrac12 t^2} \int_{-\infty}^{\infty} e^{-\tfrac12 (t - \sqrt{n}\mu / \sigma)^2}  \frac{d\mu}{|\mu|}\\
       = {}& c e^{\tfrac12 t^2} \int_{-\infty}^{\infty} e^{-\tfrac12 (t - y)^2}  \frac{d\!y}{|y|},
\end{align}
where we made a change of variable $y = \sqrt{n}\mu / \sigma$ to eliminate $n$. This is completely independent of the sample size, $n$. It depends, however, on the arbitrary coefficient $c$ in our improper prior in \refeq{eq:log_prior} and the integral diverges. A dependence on scale must be reintroduced when making the prior proper, as we must modify the form of the prior so that it diverges slower than $1 / |\mu|$ as $\mu \to 0$ and tends to zero faster than $1 / |\mu|$ as $|\mu| \to \infty$. The dependence may, however, be weak.

For example, suppose we give support to a scale-invariant prior on only a finite region,
\begin{equation}\label{eq:proper}
p(\mu) = \begin{cases}
         \frac{1}{2\log(b / a)}\frac{1}{|\mu|} & a \le |\mu| \le b\\
         0 & \text{elsewhere},
         \end{cases}
\end{equation}
where $0 < a < b$. This prior is proper. The resulting Bayes factor may be written as
\begin{equation}\label{eq:b10_proper}
B_{10} = \frac{e^{\tfrac12 t^2}}{2\log(b / a)} \int\limits_{\sqrt{n} a / \sigma}^{\sqrt{n} b / \sigma} \left[e^{-\tfrac12 (t - y)^2} + e^{-\tfrac12 (t + y)^2}\right] \frac{d\!y}{|y|},
\end{equation}
which depends on the sample size only through the integration limits. The integral converges and the Bayes factor is bounded by $e^{\frac12 t^2}$. The factor in the square brackets peaks at or near $y = |t|$. So long as that peak lies within the integration limits, the dependence on the integration limits, and thus on the sample size, remains weak.

\begin{figure}[t]
\centering
\includegraphics[width=0.95\linewidth]{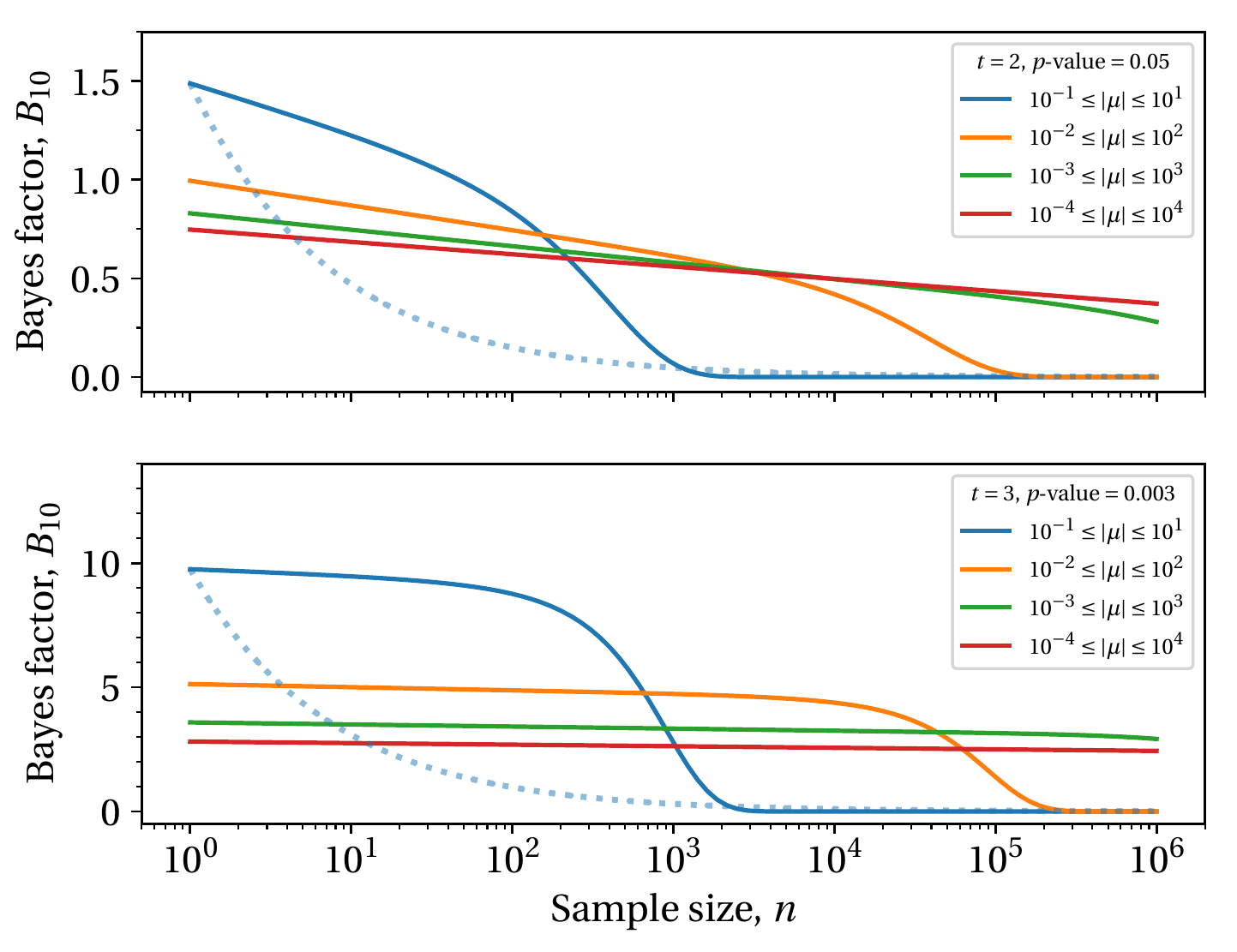}
\caption{The Bayes factor in \refeq{eq:b10} as a function of sample size $n$ for $\sigma=1$ and fixed $t = 2$ (top) and $3$ (bottom). We show Bayes factors from five different breadths of scale-invariant prior by different colours. We show $1/\sqrt{n}$ scaling by a dashed line.}\label{fig:b10}
\end{figure}

Indeed, we show the Bayes factor in \refeq{eq:b10_proper} as a function of $n$ for fixed $t = 2$ and $3$, and $\sigma = 1$ in \reffig{fig:b10}. We show results from five priors of type \refeq{eq:proper} with different choices of $a$ and $b$ by different colours, e.g., from $10^{-1} \le |\mu| \le 10^1$ (blue).  We see that as we increase the breadth of the prior, the Bayes factor remains constant as $n$ increases for greater and greater $n$, especially when $t = 3$. Once the sample size increases such that the $\mu$ preferred by the likelihood at fixed $t$ lies outside or close to the edge of the region supported by the prior, however, the Bayes factor tends to zero. The JL paradox is thus delayed until the likelihood favours $\mu$ forbidden by the prior. This delay is practically significant if we are concerned with realistic experiments in which $n$ cannot be increased indefinitely.

As emphasised by \refcite{10.2307/2332888}, the Bayes factor is sensitive to the diffuseness of the prior, e.g., the normalising constant $\log(b / a)$ required in \refeq{eq:proper}. In this problem, however, the Bayes factor may remain finite even as $\log(b / a)$ diverges, depending how the limit is taken. Let us momentarily take $t > 0$ for simplicity. If we consider,
\begin{equation}
p(\mu) = \begin{cases}
         \frac{1}{4c} \frac{1}{|\mu|} & e^{-c} \le |\mu| \le e^c\\
         0 & \text{elsewhere},
         \end{cases}
\end{equation}
and take $c \to \infty$, we asymptotically put a quarter of the prior mass between $\mu = 0$ and the maximum likelihood $\mu = t / \sqrt{n}$ for fixed $t$. Of the other three quarters of the prior mass, half lies at the wrong sign of $\mu$ and another quarter lies at $\mu > t / \sqrt{n}$. In the region $0 \le \mu \le t / \sqrt{n}$, the likelihood is always greater than that for $\mu = 0$ and thus there is a lower bound on the Bayes factor of one quarter. Thus Bartlett's paradox is not particularly severe in this case. In practice the limits $a$ and $b$ in \refeq{eq:proper} should be set by considering realistic maximum and minimum scales for the unknown mean in a concrete problem with real information about the meaning of $\mu$.

\section{Discussion}\label{sec:conclusions}

In the JL paradox, we consider a fixed $t$-statistic but a changing sample size, $n$. Fixing $t$ but changing $n$ is arguably a change in prior knowledge about the origin of $t$, since if we were merely collecting more samples, fixing $t$ would be unrealistic~\pcite{doi:10.1086/675729}. We were thus motivated to construct priors that are completely indifferent to $n$ because it puts identical information into the Bayesian calculations and the \pvalue, as the latter is independent of $n$. Since the JL paradox concerns an observation of $t$ and an unknown mean, we desired a prior for the mean that was independent of $n$ and that resulted in a prior predictive for $t$ that was independent of $n$. It is self-evident that this would result in a Bayes factor that is independent of $n$. Thus, for such a prior, the JL paradox would vanish, since we could no longer exacerbate differences between results from Bayesian and frequentist tests by increasing $n$.

We showed that the prior representing indifference to $n$ in this problem is a scale-invariant prior for the unknown mean $\mu$. This isn't surprising: by considering a fixed test-statistic but increasing sample size, we are in fact just re-scaling the problem, and by using a scale-invariant prior, we are indifferent to that change. This prior is, however, improper, and making it proper necessarily re-introduces a dependence on scale. The dependence on scale may, however, be weak until $n$ is increased so much that at fixed $t$ it favours $\mu$ that is so small that it lies in the region in which the scale-invariant prior was modified. Since a proper prior must diverge slower than $1/|\mu|$ at small $\mu$, it must disfavour (rather than favour) smaller scales relative to the scale-invariant prior, and thus favour the null model when $n$ is sufficiently increased.

We see, then, that a JL paradox is inevitable at sufficiently large $n$, because there is no scale-invariant proper prior over the entire real line.  Thus, for fixed $t$, not only is it inevitable that the Bayesian result ultimately depends on sample size, it is also inevitable that it ultimately favours the null model. In this light, the JL paradox is no longer troubling: the results from a Bayesian analysis are only independent of the sample size if we had a specific prior state of knowledge that was indifferent to the sample size. Such a state of knowledge is not entirely possible, as the resulting scale-invariant prior is improper. For any other choice of prior, the result depends on $n$ and thus we find a JL paradox when we compare against the frequentist result, which is automatically independent of the sample size. We showed, however, that we may modify the scale-invariant prior to make it proper and delay the strong dependence on sample size.

Lastly, we can think about this another way. If the Bayesian answer was troubling, we should carefully think about what question we were asking~\citep{Bernardo1980}. At first glance, we might think we were asking whether $\mu = 0$ or $\mu \neq 0$ --- this is the question that the frequentist approach appears to answer. A proper prior, however, cannot be scale invariant; at the least, it must disfavour arbitrarily tiny and enormous scales. Thus, we are really asking, is $\mu = 0$ or is it the scale favored by the choice of prior? Asked this way, it is intuitive that for fixed $t$ the preference for $\mu = 0$ must eventually grow as $n$ increases, as eventually the data favors $\hat\mu = t \sigma / \sqrt{n}$ smaller than that favored by the prior.

\bibliographystyle{ba}
\bibliography{references}
\end{document}